# Chip-scale, CMOS-compatible, high energy passively Q-switched laser

Neetesh Singh[1*], Jan Lorenzen[1,2], Milan Sinobad[1], Kai Wang[3], Andreas C. Liapis[4], Henry Frankis[5], Stefanie Haugg[6], Henry Francis[7], Jose Carreira[7], Michael Geiselmann[7], Mahmoud A. Gaafar[1], Tobias Herr[1], Jonathan D. B. Bradley[5], Zhipei Sun[4], Sonia M Garcia-Blanco[3], and Franz X. Kärtner[1,8]

[1]*Center for Free-Electron Laser Science CFEL, Deutsches Elektronen-Synchrotron DESY, Germany*
[2]*University of Kiel, Kiel, Germany*
[3]*Integrated Optical Systems, MESA+ Institute for Nanotechnology, University of Twente, 7500AE, Enschede, The Netherlands*
[4]*Department of Electronic and Nanoengineering, Aalto University, Espoo 05140, Finland*
[5]*Department of Engineering Physics, McMaster University, 1280 Main Street West, Hamilton, Ontario L8S 4L7, Canada*
[6]*Centre for hybrid Nanostructures, University of Hamburg, Luruper Chaussee 149,, 22761, Hamburg, Germany*
[7]*LIGENTEC SA, EPFL Innovation Par L, Chemin de la Dent-d'Oche 1B, Switzerland CH-1024 Ecublens, Switzerland.*
[8]*Department of Physics, Universität Hamburg, Jungiusstr. 9, 20355 Hamburg, Germany*
*\*neetesh.singh@desy.de*

**Abstract:** Chip-scale, high-energy optical pulse generation is becoming increasingly important as we expand activities into hard to reach areas such as space and deep ocean. Q-switching of the laser cavity is the best known technique for generating high-energy pulses, and typically such systems are in the realm of large bench-top solid-state lasers and fiber lasers, especially in the long wavelength range >1.8 µm, thanks to their large energy storage capacity. However, in integrated photonics, the very property of tight mode confinement, that enables a small form factor, becomes an impediment to high energy application due to small optical mode cross-section. In this work, we demonstrate complementary metal-oxide-semiconductor (CMOS) compatible, rare-earth gain based large mode area (LMA) passively Q-switched laser in a compact footprint. We demonstrate high on-chip output pulse energy of >150 nJ in single transverse fundamental mode in the eye-safe window (1.9 µm), with a slope efficiency ~ 40% in a footprint of ~9 mm$^2$. The high energy pulse generation demonstrated in this work is comparable or in many cases exceeds Q-switched fiber lasers. This bodes well for field applications in medicine and space.

**Introduction:** Soon after the invention of the laser, the '*giant pulse*' laser was experimentally demonstrated by McClung and Hellwarth [1]. Ever since its invention, Q-switching has remained the common technique for generating pulses of very high energy which found applications in range finding and sensing, micromachining, and medicine among others [2-11]. In the eye safe wavelength in the extended-short-wave infrared >1.7 µm, which is important for biological imaging, environmental monitoring, spectroscopy, free space communications and even recently envisaged local area networks (where fiber loss is acceptable) [12-15], a high energy pulsed laser can play a key role. For example, Q switched lasers are used in laser surgery for precise cutting of biological tissue, such as in ophthalmic and spinal surgery [5-8]. They are used in the food, semiconductor and electronics industries for humidity control, and in differential absorption LIDAR for mapping wind, water and carbon dioxide in the atmosphere [5, 9-11].

Currently, the industry is dominated by large solid-state and fiber lasers that provide the desired power and energy, thanks to their large energy storage capacity owing to large signal mode area and long cavity length (mode area *x* length). Here, large mode area is beneficial because, a) it reduces the instability caused by optical nonlinearities due to high intensity, and more importantly, b) it helps to increase the signal power or energy by increasing the gain saturation power which is given as, $P_{sat}=E_{sat}/t_l$, where $t_l$ is the upperstate lifetime and the $E_{sat}$ is the saturation energy of the gain medium which is proportional to the mode area, $E_{sat} \propto A_{eff}$. In other words, the larger volume increases the stored energy. Therefore, a high-power fiber laser typically employs LMA fiber to reach power and energy levels usually enjoyed by solid state lasers [16, 17]. Achieving such a capability on a chip-scale will have a disruptive impact in the point-of-care and field-deployable systems such as space-bound instruments for earth and planetary LIDARs where size, weight and power (SWaP) is critical [9-11]. In integrated photonics, however, the energy storage capacity is severely limited compared to its benchtop counterpart. This is because the cavity length and the mode area are usually limited to around a few cms (with a reasonable loss) and a few µm$^2$, respectively. Recently this limitation has been somewhat relaxed in one-dimension (length) by the fabrication of low-loss long (0.5 m) silicon nitride (SiN) waveguides allowing high signal amplification [18], but at the cost of a complex fabrication process. On the other hand, semiconductor gain media can achieve very high gain over a short length due to their large emission cross-section [19, 20]. However, semiconductor gain media face challenges with short upperstate lifetime, high thermal instability and high nonlinear losses such as two photon absorption. Nevertheless, progress has been made with CW lasers such as those based on gallium antimonide operating around 1.9 µm, producing (albeit) a few milliwatts of power [21, 22]; moreover, large mode area structures (currently for 1.55 µm) are also being investigated for integration with silicon photonics [23, 24].

In this work, we demonstrate CMOS compatible large mode area waveguides [25] that support several tens of µm$^2$ mode area and long cavities within a compact footprint [26, 27], and have demonstrated Q-switched high energy laser with fiber laser-level performance. Our LMA design is based on a single thick silicon nitride layer that allows tighter bends for compact footprint (which also allows seamless integration of conventional nonlinear photonics components [28]). The thickness of the SiN can vary depending on wavelength and foundry process which simplifies fabrication and allows adaptation to different CMOS foundries (unlike previous designs with multiple SiN layers that support a small mode area and higher order modes [29-31]). Our LMA design, unlike an LMA fiber laser, supports only fundamental mode propagation and allows high overlap of pump and signal modes even when they are spectrally far apart [25-27]. Since the optical modes are mainly guided in the gain film, the gain waveguide propagation loss is mainly determined by the film loss which is intrinsically lower than that of an etched waveguide [32, 33]. In this work, using the LMA waveguide and rare earth gain medium (now used in integrated photonics [34-46]), we have demonstrated a Q-switched high energy laser achieving pulse energies over 150 nJ around 1.9 µm within a small device footprint of ~ 9 mm$^2$. The Q switched pulse energy demonstrated here is ~20 dB higher and the slope efficiency is ~ 40% as compared to 1% in the previous demonstration [29]. More importantly, the slope efficiency is higher, and the pulse energy is comparable or in many cases higher than many benchtop Q-switched all-fiber lasers using true saturable absorbers [47-55]. Moreover, higher peak power with shorter pulses can be achieved by tuning the saturable absorber, provided that we operate within the optical damage threshold of the SiN waveguide as one can reach intracavity pulse energy close to the damage threshold of SiN waveguides with nanosecond pulses [56-58].

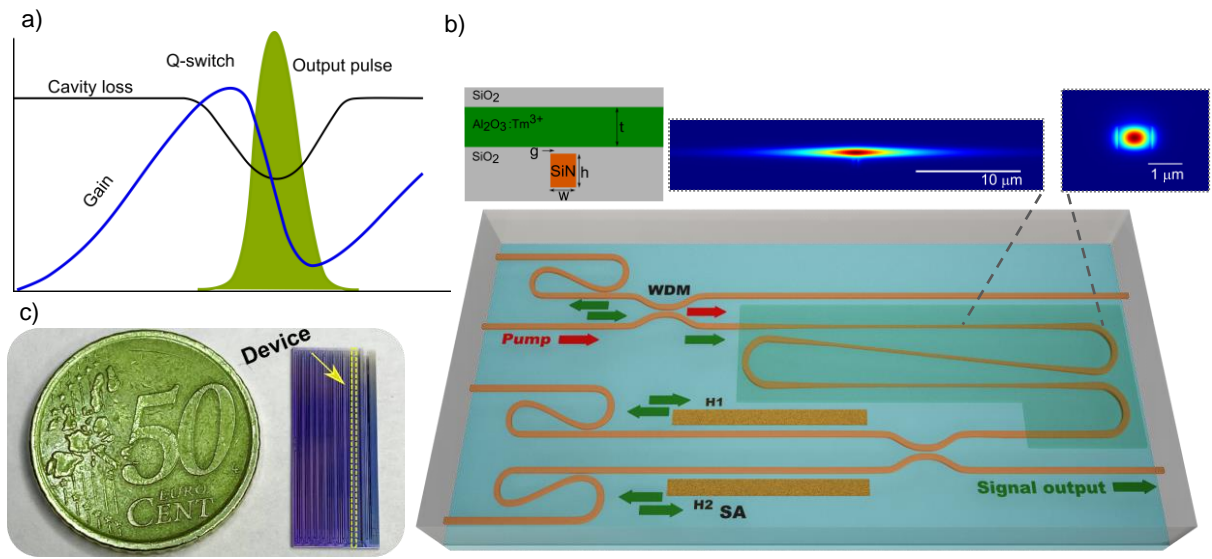

*Figure 1. a) Principle of Q-switching. b) A compact integrated LMA pulsed laser. WDM – wavelength division multiplexer; SA – saturable absorber; H1,2 – high and low power heater arms. The pump is coupled into the gain section through a WDM towards the serpentine waveguide. Within the green box region, the LMA sections are over the straight SiN waveguides. The integrated SA is a nonlinear Michelson interferometer (NLI) which splits the signal into two arms through a directional coupler and backreflects via loop mirrors. Inset: LMA waveguide cross-section is shown along with signal mode profiles at the LMA gain region and at a compact bend. c) An optical image of the chip containing the laser (dashed region on the chip indicated by an arrow) against a 50 cent euro coin.*

**Integrated Q-Switched laser:** For decades, Q-switching has been a technique to generate highly energetic pulses from a laser by modulating the loss and thus the quality factor (Q) of the cavity. When the Q of the cavity is low, that is when the stimulated emission is suppressed, the pump excites the gain ions to an upper state storing energy until a high population inversion is achieved. Sudden change of Q to a high state (by reducing the loss) allows stimulated emission to take place causing an instantaneous conversion of stored energy in the gain medium into signal photons and a rapid buildup of a high power circulating signal. In doing so, the photons deplete the gain causing signal amplification to stop as soon as the gain drops below the cavity loss, which results in a giant Q-switched pulse emission. A cartoon showing this process is shown in Fig. 1a. Switching of the cavity Q can be attained actively or passively. Active Q-switching requires an externally controlled intracavity modulator such an electro-optic device, which makes the system complex and bulky. Passive Q-switching on the other hand, requires a saturable absorber whose loss drops with pulse intensity, which is a cost effective, compact and robust technique. For this reason, there is a strong desire in the fiber community to develop passively Q-switched high energy lasers, and several groups have tried different types of saturable absorbers over the past decade that are durable, inexpensive, have a high damage threshold, and require little space [47-55, 59-63].

In this work, we demonstrate an integrated passively Q switched laser utilizing an artificial saturable absorber that is durable and has high damage threshold limited only by the damage threshold of the SiN waveguide [25, 30, 56]. The laser mainly consists of a gain section, cavity mirrors and a saturable absorber (Fig. 1b). Here, the pump is coupled to the gain section through an integrated wavelength division multiplexer (WDM), which guides the pump to the gain medium and in reverse the signal to an integrated loop mirror reflector. The pump is then adiabatically coupled into the gain medium, which is based on thulium doped aluminum oxide film (~820 nm thick) which is cladded with 1 μm thick PECVD silica layer. The cross-section of the LMA gain waveguide is shown along with the simulated fundamental TE mode profile of the signal in Fig. 1b (inset). The signal mode area = 28 μm$^2$ and the pump mode area = 27 μm$^2$ giving a pump-signal overlap of >99%. The thickness of the silicon nitride (SiN) layer in the gain section is 800 nm (h), interlayer oxide thickness = (g) 300 nm and the width (w) = 290 nm. The TE mode power of the signal is ~75% in the aluminum oxide gain film, 0.2% in the SiN layer and 11.8% and 13% in the top and the bottom silica layer, respectively. A narrow SiN layer not only reduces scattering loss (due to a weak modal confinement) but also helps to maintain only the fundamental mode in the gain section. The pump and signal can be coupled efficiently into the large fundamental TE mode in the gain with adiabatic tapers, the tapers also help to pull the large modes back into the SiN layer to allow tighter bends [25]. To obtain a long cavity while keeping the footprint small, we use a serpentine shaped gain waveguide with circular bends as shown in Fig. 1b (3D). Near the output end of the gain section the signal is adiabatically coupled into a wide SiN waveguide, which leads to an integrated saturable absorber. We use a fast Kerr-type saturable absorber based on nonlinear Michelson interferometer (NLI-SA) [25, 30]. It acts as an intensity dependent reflector - the reflectivity back to the cavity increases as the signal intensity increases, favoring a shorter pulse in the cavity, which supports pulse formation, for details see [25]. The NLI-SA consists of a non-3dB coupler and two symmetric arms terminated with loop mirrors (or gratings in other configurations). The non-3dB coupler splits the signal into two arms with different power, which, by propagating in nonlinear waveguides, acquire a power dependent Kerr phase shift difference with respect to each other that increases with signal power (a π phase difference, for example, allows for full signal reflection back to the cavity).

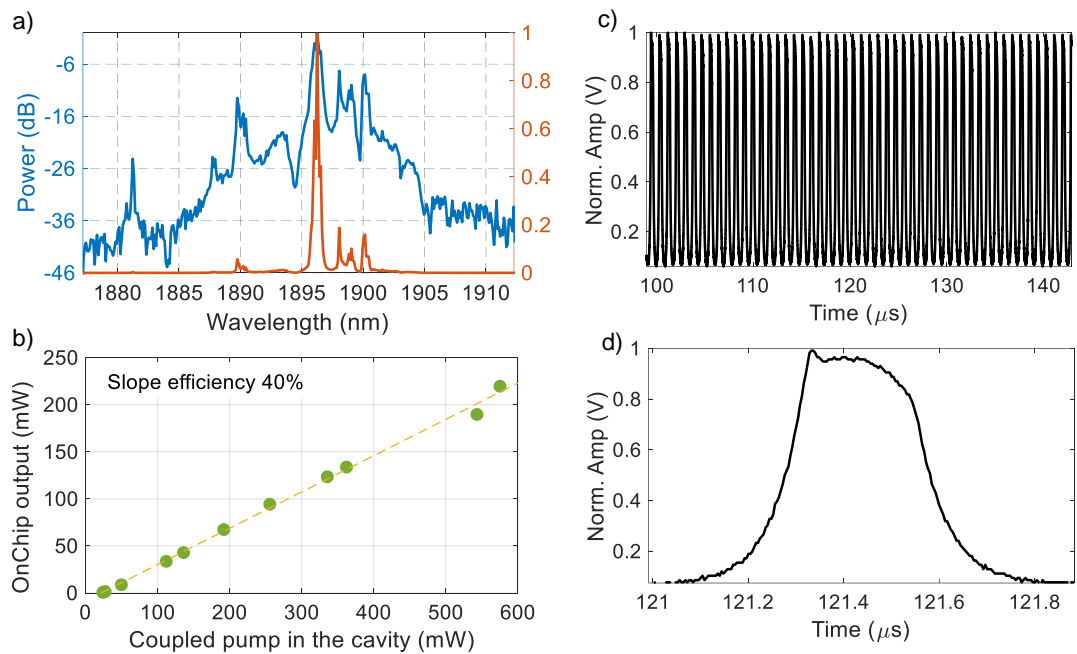

*Figure 2. a) The optical spectrum in logarithmic and linear scale. b) The coupled pump and on-chip output average signal power. c) The pulse train of the Q-switched signal, and d) a zoomed in section showing a pulse*

**Experiment and Results:** The passive layer of the device was fabricated in a silicon photonics foundry (*Ligentec*). A thick SiN layer was deposited, polished, patterned, and subsequently cladded with a silica layer. A local oxide opening is created where the silica layer is removed from the top of the SiN region down to the point where it's only ~ 300 nm (g) above the the SiN layer and a gain layer is deposited (green box Fig.1b). For the gain deposition, a back end of the line 820 nm thick thulium-doped alumina gain layer (Tm$^{3+}$:Al$_2$O$_3$) is deposited by RF sputtering with an estimated thulium ion concentration of 3.2x10$^{20}$cm$^{-3}$. The passive film loss is measured separately with the prism coupling technique (*Metricon)* to be 0.83dB/cm @ 450nm, and 0.55dB/cm @ 520nm which corresponds to a loss <0.1dB/cm beyond 1.55 μm as measured in ref. [32, 33]. The refractive index of the film was estimated at 1.9 μm to be 1.70 using an ellipsometer with Sellmeier fitting. In the gain section, the bend loss is measured to be less than -0.0008 dB/180 degree of a bend radius of 90 μm.

The broadband spectral response of the passive components is measured with a supercontinuum source (*NKT*). The WDM has a bandwidth of 130 nm (@ 1 dB level) around the signal at 1.89 μm, and the pump port (at 1.61 μm) has >90% transmission. The measured 3dB crossing point of the cross port and through port of the directional coupler of the loop mirror is centered ~ 1.89 μm, giving an estimated to reflection bandwidth of 170 nm (@ 1 dB level). The power splitter (a directional coupler) of the NLI-SA (saturable absorber) has a split ratio of 80:20 at 1.9 μm resulting in modulation depth of >50%. Heaters are integrated to compensate for fabrication uncertainties, i.e. to properly *bias* the NLI-SA (i.e. to compensate for the phase offset between the two arms caused by fabrication tolerances), and to vary the self-amplitude-modulation parameter of the NLI-SA. The NLI-SA bias is the key parameter to control the mode of operation of the laser (i.e. pulsed or CW) as we explain later in more detail.

In the experiment, the pump at a wavelength of 1.61 μm was launched into the chip with a lensed fiber. The laser starts to operate in CW-mode, and as the pump power increases inherent relaxation oscillations build up and passively Q-switched high energy pulses emerge, which is due to the presence of a loss modulator such as a saturable absorber as mentioned above. The optical spectrum of the signal measured at 400 mW of coupled pump power in the cavity is shown in Fig.2a which is measured at the output port. The measured slope efficiency of the laser is 40% and the lasing threshold pump power is around 20 mW (Fig. 2b). The pulses are produced at <1 MHz repetition rate with a pulse width of 250 ns and an energy over 150 nJ. The signal is centered at 1.89 μm which is also the maximum transmission and reflection wavelength of the pump and signal combiner (WDM) and the loop mirror, respectively. One can apply a narrow bandpass filter at the output or co-integrate a narrowband filter to extract the narrowband signal around 1896 nm, which dominates the signal energy (as seen in Fig 2a, linear plot)).

As mentioned earlier, due to fabrication uncertainties the NLI-SA reflectivity is tuned with an integrated phase shifter to ensure that the signal lies on the positive reflectivity slope of the NLI-SA to allow for pulse formation with high energy (as in Q-switching). The ideal reflectivity curve of an NLI-SA is shown in Fig. 3a. Here Φ(p) is the input peak power dependent nonlinear differential phase shift of the light in the two interferometer arms. With a non-3dB splitter the two arms have different signal power, therefore, Φ(p) and the reflectivity back to the cavity increase with signal power. The modulation depth, the magnitude of the reflectivity slope (self-amplitude-modulation coefficient) and the signal power required to reach the peak reflection increase as the power splitting ratio of the coupler increases [25]. In the positive slope (increasing reflectivity) in Fig. 3a, the laser cavity tends to form pulses such as through Q-switching, Q-switch mode locking and CW mode locking and in the negative slope (decreasing reflectivity) pulse formation is suppressed and the laser operates in the CW-mode.

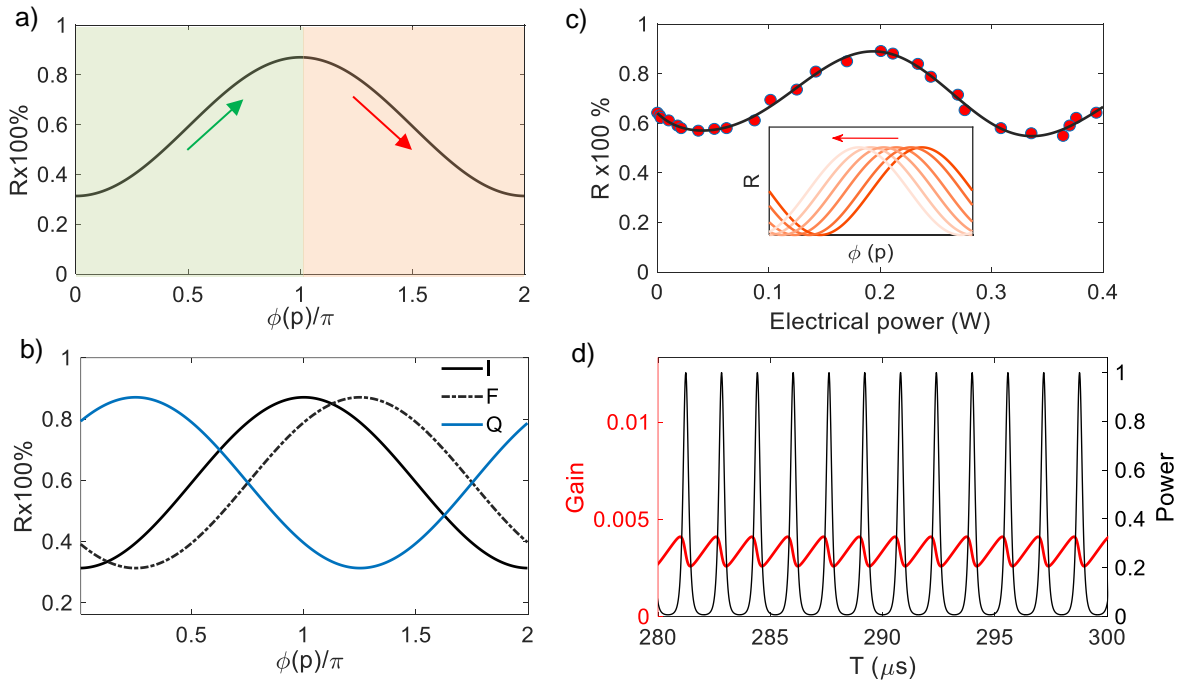

*Figure 3. a) An ideal NLI-SA reflectivity curve as a function of power dependent differential phase shift. Green and red arrows indicate pulsed operation region (positive slope) and CW operation region (negative slope), respectively. b) The NLI reflectivity curve of an ideal (I), fabricated (F) and the case where the curve is tuned for Q-switching (Q). c) Measurement of thermally tuned NLI reflectivity variation of coupled input signal of an optical power equivalent to Φ(p)~0 to help extract the overall curve. The inset shows a cartoon of how the reflection strength varies (for a signal with Φ(p)~0) due to the movement of the curve with thermal phase shift which helps to trace the as-fabricated NLI-SA reflectivity curve, Fig. 4b (dashed curve). d) Simulated time domain response of the laser showing a Q-switched pulse train and variation in gain.*

With an integrated phase shifter, the NLI-SA can be tuned to switch the laser between the pulsed and the CW-lasing mode. This may find applications in laser surgery where the two modes of operation are used, such as CW-laser for neurosurgery and pulsed laser for spinal surgery [5].

The as-fabricated NLI-SA, however, doesn't have an ideal behavior due to fabrication uncertainties, causing the reflectivity curve to be shifted (Fig. 3b). One can extract the reflectivity response of the as-fabricated NLI-SA in a laser as shown in Fig. 3c. In that, we couple a very weak test signal (supercontinuum NKT source) in the NLI-SA such that $\Phi(p) \sim 0$ and then gradually increase the thermal phase shift (with an integrated heater), which varies the reflectivity curve, and collect the reflected light which varies in power (due to the change in reflectivity). As shown in Fig. 4c (inset), the reflectivity curve is tuned to the left and the reflected power measured near ($\Phi(p) \sim 0$) varies accordingly - first it drops to a minimum and then increases (in an ideal case it should start by increasing from the minimum). Using this method, we trace out the response of the fabricated NLI-SA, as shown in Fig. 3b. We note that the peak to peak difference (in Fig. 3b) is < 40% instead of being larger as expected for a DC with 80:20 splitting ratio at 1.9 µm (Fig.2c), this is mainly due to using a broadband source for testing causing higher signal reflection (from unwanted spectral bands) than is expected from a narrow band source around 1.89 µm. Once the NLI curve is known, it can be shifted either to the right or to the left against the horizontal axis with the heater tuning of the two arms so that the lasing signal falls on the positive slope of the curve. To obtain Q-switched high energy pulses we shifted the curve to the right (blue curve (Q) in Fig. 4b) until the reflectivity back to the cavity is increased for higher intracavity power and high energy pulses formed (on the positive NLI-SA slope), which happened around $\Phi(p) \sim 0.1$, corresponding to an intracavity pulse peak power of ~ 6-10W. The curve can also be shifted to the left such that the signal falls near the bottom of the positive slope, however, in that case the reflectivity is low thus only low energy pulses can be extracted. We note, for every pump power the reflectivity curve of the NLI-SA is slightly readjusted with the heaters such that the signal falls close to the peak reflectivity in the positive slope region. For high power CW operation, the NLI can be tuned such that signal lies on the negative slope near the peak of the reflectivity. The thermal cross talk between the two arms of the NLI is expected to be negligible as they were separated by 30 µm. We simulated the behavior of the Q-switched laser by solving coupled rate equations [64, 65] with a modified saturable absorber loss function (*cosine* function to account for the NLI-SA). Since we are operating near the very peak on the positive slope of the NLI reflectivity curve the effective saturable absorber's modulation depth and the saturation power of the saturable absorber is low. This results in a very weak gain modulation (well below a percent), as shown in Fig. 3d, causing long pulse formation.

We monitored the stability of the output power of the Q-switched signal over 25 minutes, as shown in Fig. 4. The fluctuation is about 1 to 2 % which is mainly due to the mechanical drift in the alignment of the chip against the lensed fibers; with realignment the signal remained stable for several hours (not shown).

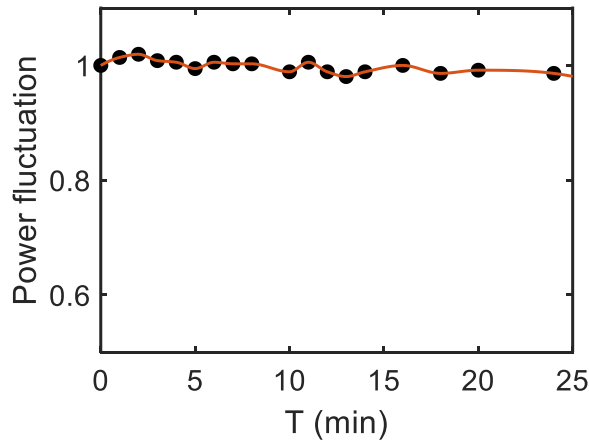

*Figure 4. The normalized output power of the Q-switched laser over a period of several minutes.*

**Conclusion and outlook:** In summary, we have demonstrated CMOS-compatible Q-switched high energy laser in a compact footprint which is comparable to, and many cases exceeds, a benchtop fiber laser. This bodes well for wide range of applications where size, weight and power are important, such as in deep space and ocean and point-of-care medical applications. Higher energy can be extracted with even larger mode area which will increase the gain saturation power over the current design, reaching mode sizes comparable to fiber lasers, and energy in the several microjoule range can be obtained. Furthermore, shorter pulses (sub-nanosecond range) with high peak power can be extracted by increasing the NLI's saturable loss and saturable power by thermal tuning while maintaining a higher energy per pulse. However, caution must be exercised in order to limit the intracavity pulse energy (which can reach over a microjoule level) to be within the damage threshold of the SiN waveguide (which scales with the pulse width as ~$\tau^{0.5}$ for $\tau$ >10 ps) [57]. Such thresholds are easy to reach in the regions where the modes are tightly confined such as in the SiN bends, which can have a damage threshold in the range of 100s of nanojoule to a microjoule (for nanosecond pulses) [56-58]. Moreover, our design is scalable to different wavelength windows, such as telecom window (1.55 µm) and high-power laser window at 1 µm (based on ytterbium dopants), where optical loss of the SiN and aluminum oxide is fairly low. The device footprint can be further reduced down to <5 mm$^2$ with Euler bends and uniform Bragg grating reflectors. One can also integrate apodised chirped grating in place of the loop mirrors to manage the cavity dispersion. This will allow one to achieve high peak power Q-switched mode locking (QSML) or CW mode locking for ultra-stable pulse generation for application requiring high peak power and low noise pulse train.

**Acknowledgements:**
This work is supported by EU Horizon 2020 Framework Programme - Grant Agreement No.: 965124 (FEMTOCHIP), and Deutsche Forschungsgemeinschaft (SP2111) contract number PACE:Ka908/10-1.

**Competing interest:**
The authors declare no competing interests.

**Methods:**
**Fabrication:** As mentioned above, the device was fabricated in a silicon photonics foundry (*Ligentec*) using stepper photolithography on a 100 mm wafer. The standard variation in SiN thickness and refractive index was ± 5%, and 0.25%, respectively. The photonic stack consists of silicon, bottom silicon dioxide, silicon nitride, top silicon dioxide and aluminum oxide. The thickness of silicon substrate ~ 230 µm, bottom oxide = 4 µm and the top oxide thickness = 3.3 µm. The sidewall angle of the SiN waveguide is 89°. After the patterning of the SiN layer an opening was created (after etching >300 µm wide silica away) on top of the SiN waveguide. The SiN layer and the opening was separated by a thin layer of silica of 300 nm thickness (designed value). Fill patterns of SiN were fabricated to maintain high enough density of SiN (>20%) across the reticle to avoid fabrication complications. The gain layer was deposited at University of Twente with an RF sputtering tool. The chip was mounted in a holder and is loaded into a AJA ATC 15000 RF reactive co-sputtering system through a load-lock and is placed on a rotating holder in the main reaction chamber. A two-inch aluminum target (99.9995% purity) and thulium target are powered through their own RF sources. A power of 200 W is used on the aluminum target and 18 W is used on the thulium target which determines the ion concentration in the film. The deposition temperature is around ~400C (measured with a thermocouple) and the rate of deposition is around 4 to 5 nm/min amounting to 3 hours of deposition. The devices are subsequently protected by a PECVD silica cladding layer, we note that a silica cladding layer can be replaced with a CYTOP layer, which is a well-established long wavelength cladding material [66]. The heaters were deposited at Aalto University with 10 nm of titanium and 400 nm of gold.

**Experiment (Q switching):** In the experiment, the device was optically pumped by an amplified low noise CW laser (Alnair labs, TLG 220). We used a high power polarization maintaining L band amplifier (IPG EAR-10-1610-LP-SF). The pump was launched through a half and quarter waveplate into a 99:1 fiber splitter followed by a lensed fibre (OZ optics, 3 µm spot size) which was followed by the chip. The pump was monitored with a 1% drop port of the splitter. The coupling to chip loss was measured to be between 2.5 dB to 3 dB at the pump wavelength, and at the signal wavelength it was slightly higher than 3 dB. The lasing signal was measured with an extended InGaAs integrating sphere photodiode (Thorlab 148C), and is subsequently measured on the OSA (Yokogawa AQ6376). The pulses were measured with an extended InGaAs 12 GHz detector (EOT ET 5000 F/APC) and an oscilloscope with a bandwidth of 300 MHz (RS pro RSDS 1304 CFL).

*NLI tuning experiment:* To determine the NLI reflectivity curve after fabrication. We coupled a very weak signal around 1.9 µm from a broadband NKT source (FIU-15). The broad band source was first filtered with a long pass filter (>1.5 µm) and then a bandpass filter with a 200 nm bandwidth (Thorlab FB 1900-200). The light was coupled into a fiber based 2x2 3-dB splitter which was coupled through a lensed fiber to the NLI of the laser under test. The heater on one of the arms of the NLI was tuned with the current ranging from 0 mA to 55 mA with the resistance varying from 125 Ω to 135 Ω amounting to a maximum electrical power of 400 mW. The back reflected signal was measured through the 3 dB splitter while care was taken to avoid collection of signal from the facet reflection, which obviously did not change in power with heater tuning. The peak of the NLI reflectivity curve (shown in Fig. 3) is dependent on the propagation loss in the NLI waveguides (1.8 µm wide) which we estimate, as an upper limit, to be 0.15 dB/cm, giving >90% reflectivity at the peak. With lower propagation loss the peak reflectivity increases further.

*Refractive index*: The optical constants were measured of an $Al_2O_3$ film with an ellipsometer having a wavelength span from 240 nm to beyond 11 µm. The refractive index at 1.9 µm ~ 1.7.

**References:**

1. F. J. McClung and R. E. Hellwarth, "Giant optical pulsation from ruby," *J. of Appl. Phys.* 33, (1962).

2. U. Sharma, et. al.,"Highly stable tunable dual wavelength Q-switched fiber laser for DIAL applications," *IEEE Photon. Tech. Lett.* 16 (2004).
3. M. Nagele et.al. "Passively Q-switched 914 nm microchip laser for lidar systems," *Opt. Express,* 29, (2021)
4. M. Skorczakowski, et. al., "Mid infrared Q switched Er:YAG laser for medical applications," *Laser Phys. Lett* 7, (2010).
5. K Scholle, et. al. "2 μm Laser Sources and Their Possible Applications," *Frontiers in guided wave optics and optoelectronics*, 2010.
6. K. Yang, et. al. "Q-Switched 2 micron solid-state laser and their applications," *Frontiers in guided wave optics and optoelectronics* (2019).
7. C. Boone, "Medical applications are a surgical fit for 2 μm lasers," *Laser Focus World*, 2022
8. X. Xie, et. al. "A brief review of 2 μm laser scalpel," *IEEE 5th Optoelectronics* 2022.
9. U. N. Singh, "Progress on high energy 2 micron solid state laser for NASA space-based wind and carbon dioxide measurments," 011 *IEEE Photonics Society Summer Topical Meeting Series*, 2011.
10. W. Yu, et. al. "Orbiting and In-Situ Lidars for Earth and Planetary Applications," *IGARSS 2020-IEEE Int. Geoscience and Remote Sensing Symp.* 2020
11. D. Guilhot et. al., "Laser technology in photonic applications for space,"*MDPI Photon. Dev. and Appl*. 3, 2019.
12. O. T. Bruns, et. al. "Next-generation *in vivo* optical imaging with short-wave infrared quantum dots," *Nature. Biomed. Engn.* 1, (2017).
13. N. Kakouta, et. al. "Near-infrared imaging of water vapour in air", *Meas. Sci. Technol.* **33** (2022).
14. J. Hodgkinson et. al. ,"Optical gas sensing: a review," *Meas. Sci. Technol*. 24, (2013)
15. Z. Li. Et. al. "Thulium-doped fiber amplifier for optical communications at 2 μm," *Opt. Express*, 21, (2013).
16. J. Nilsson and B. Jaskorzynska, "Modeling and optimization of low-repetition-rate high-energy pulse amplification in cw-pumped erbium-doped fiber amplifiers," *Opt. Lett.* **18**(24), 2099 (1993).
17. D. J. Richardson, et. al. "High power fiber lasers: current status and future perspectives," *J. Opt. Soc. Am. B* **27**(11), B63–B92 (2010).
18. Y. Liu, et. al. "A photonic integrated circuit based erbium doped amplifier," *Science*, 376, 2022.
19. J. Leuthold et. al. "Material gain of bulk 1.55 mm InGaAsP/InP semiconductor optical amplifiers approximated by a polynomial model," *J. of Appl. Phys*, 87, (2000).
20. D. Geskus, et. al. "Giant Optical Gain in a Rare-Earth-Ion-Doped Microstructure," *Adv. Mater.* 24 (2012) .
21. R. Wang, et. al. "Compact GaSb/silicon-on-insulator 2.0x μm widely tunable external cavity lasers", *Opt. Express*, 24 (2015).
22. N. Zia et. al. "Hybrid silicon photonics DBR laser based on flip-chip integration of GaSb amplifiers and μm-scale SOI waveguides," *Opt. Express*, 30, 2022.
23. H. Zhao, et. al. "High power indium phosphide photonic integrated circuits," *IEEE J Sel Top Quantum Electron* 25, (2019).
24. P. W. Juodawlkis et. al. "High-Power, Low-Noise 1.5-μm Slab-Coupled Optical Waveguide (SCOW) Emitters: Physics, Devices, and Applications," *IEEE J. Sel. Top. Quantum*, 17, 2011.
25. N. Singh et.al. "Towards CW modelocked laser on chip – a large mode area and NLI for stretched pulse mode locking", *Opt. Express*, 28, 15 (2020).
26. N. Singh et. al. "Large mode area waveguide for silicon photonics and modelocked lasers," *CLEO JTh3A*.56, 2022.
27. N. Singh et. al. "Towards CMOS compatible high power mode-locked lasers and frequency combs," *CLEO SF2G*.1, 2022.
28. V. Brasch et. al. "Photonic chip–based optical frequency comb using soliton Cherenkov radiation,", *Science*, 351 (2015).
29. K. Shtyrkova, "Fully integrated CMOS-compatible mode-locked lasers," *Thesis MIT* (2018).
30. K. Shtyrkova, et al., "Integrated CMOS-compatible Q-switched mode-locked lasers at 1900 nm with an on-chop artificial saturable absorber", *Optics Express*, vol. 27, no. 3, pp. 3542-3556, 2019.
31. F. X. Kärtner et.al.,"Integrated CMOS-Compatible Mode-Locked Lasers and Their Optoelectronic Applications", *Proc. SPIE 10686*, 14 (2018).
32. K. Worhoff, et. al., "Reliable Low-Cost Fabrication of Low-Loss Al2O3:Er3+ Waveguides With 5.4-dB Optical Gain," *IEEE J. Sel. Top. Quantum*, 45, 2009.
33. Purnawirman, et. al. "C- and L-band erbium-doped waveguide lasers with wafer-scale silicon nitride cavities," *Opt. Lett*. 38, (2013).
34. J. Kenyon, "Erbium in silicon", *Semicond. Sci. Technol.*, vol. 20, pp. R65-R84, 2005.
35. L. Agazzi, et. al., "Monolithic integration of erbium-doped amplifiers with silicon-on-insulator waveguides," *Opt. Express* **18**(26), 27703–27711 (2010)
36. M. Belt and D. J. Blumenthal, "High temperature operation of an integrated erbium- doped DBR laser on an ultra-low-loss Si3N4 platform," *Optical Fiber Communication Conference*, OSA Technical Digest (online) (Optical Society of America, 2015), paper Tu2C.7
37. E. S. Magden, N. Li, J. Purnawirman, D. B. Bradley, N. Singh, A. Ruocco, et al., "Monolithically-integrated distributed feedback laser compatible with CMOS processing", *Opt. Express*, vol. 25, no. 15, pp. 18058-18065, 2017.
38. N. Li, D. Vermeulen, Z. Su, E. S. Magden, M. Xin, N. Singh, et al., "Monolithically integrated erbium-doped tunable laser on a CMOS-compatible silicon photonics platform", *Opt. Express*, vol. 26, pp. 16200-11, 2018.
39. J. Rönn, W. Zhang, A. Autere, X. Leroux, L. Pakarinen, C. Alonso-Ramos, et al., "Ultra-high on-chip optical gain in erbium-based hybrid slot waveguides", *Nat. Communications*, vol. 10, pp. 432, 2019.
40. H. Sun, L. Yin, Z. Liu, Y. Zheng, F. Fan, S. Zhao, et al., "Giant optical gain in a single-crystal erbium chloride silicate nanowire", *Nat. Photonics*, vol. 11, pp. 589-593, 2017.

41. A Choudhary, et al., "A diode-pumped 1.5 µm waveguide laser mode-locked at 6.8 GHz by a quantum dot SESAM", *Laser Phys. Lett.*, vol. 10, pp. 1-4, 2013.
42. D. Choudhury, et. al., "Ultrafast laser inscription: perspectives on future integrated applications", *Laser Photonics Rev.*, vol. 8, pp. 827-846, 2014.
43. H. Byun, et al., "Integrated low-jitter 400-MHz femtosecond waveguide laser", *IEEE Photo. Tech. Lett*, vol. 21, pp. 763-765, 2009.
44. J. D. Bradley, et al., "Monolithic erbium- and ytterbium doped microring lasers on silicon chips", *Opt. Express*, vol. 22, pp. 12226-12237, 2014.
45. K. van Dalfsen et. al. "Thulium channel waveguide laser with 1.6 W of output power and ~80% slope efficiency," *Opt. Lett.* 39, (2014).
46. C. I. Emmerik, et. al. "Single-layer active-passive $Al_2O_3$ photonic integration platform," *Opt. Material. Express.* 10, (2020).
47. H. Ahmas, et. al. "2.08 µm Q switched holmium fiber laser using niobium carbide – polyvinyl alcohol (Nb2C-PVA) as a saturable absorber," *Optics. Commun.* 490, 2021.
48. H. Ahmad, et. al. "Generation of Q-switched Pulses in Thulium-doped and Thulium/Holmium-co-doped Fiber Lasers using MAX phase (Ti3AlC2)," *Scientific Rep.* 10, 2020.
49. M. Wang, et. al. "Passively Q switched thulium-doped fiber laser based on oxygen vacaency MoO3-x saturable absorber," *Opt. Mat. Express*, 9 (2019)
50. M. F. A. Rahman, et. al. "Q-switched and mode-locked thulium-doped fiber laser with pure Antimony film Saturable absorber," *Optics. Communication,* 421, (2018).
51. A. A. Latiff, et. al. "A generation of 2 µm Q-switched thulium-doped fibre laser based on anatase titanium(IV) oxide film saturable absorber," *J. of Mod. Optics*, (2016)
52. R. I . Woodward, et. al. "Wideband saturable absorption in few-layer molybdenum diselenide (MoSe2) for Q-switching Yb-, Er- and Tm-doped fiber lasers," *Opt. Express,* 23, (2015).
53. M. Jung, et. al. "An all fiberized, 1.89-µm Q-switched laser employing carbon nanotube evanescent field interaction," *Laser Phys. Lett,* 9 (2012)
54. M. Chernysheva, et. al. "High Power Q-Switched Thulium Doped Fibre Laser using Carbon Nanotube Polymer Composite Saturable Absorber," *Sci. Report*, 6, (2016)
55. J. Liu et. al. "Graphene-based passively Q-switched 2 µm thulium-doped fiber laser," 285, *Optics. Commun.* (2012)
56. S. Tan, et. al. "Silicon nitride waveguide as a power delivery component for on-chip dielectric laser accelerators," *Opt. Lett.* 44, (2019)
57. B. C. Stuart et. al. "Nanosecond to femtosecond laser-induced breakdown in dielectrics," *Phys. Rev. B*, 53, 1996.
58. V. Gruzdev, "Self-focusing and nonliear effects," Chpt 4, *Laser Induced damaged in optical materials*, 2014.
59. D. Popa, et. al., "Graphene Q-switched, tunable fiber laser", *Appl. Phys. Lett,* 98 (2011)
60. B. Fou, et. al. "Passively Q-switched Yb-doped all-fiber laser based on Ag nanoplates as saturable absorber," *Nanophotonics,* 9, (2020)
61. S. A. Hussain, "Discovery of Several New Families of Saturable Absorbers for Ultrashort Pulsed Laser Systems," *Sci. Report.* 9, (2019)
62. J. Lee, et. al. "Ti2AlC-based saturable absorber for passive Q-switching of a fiber laser", *Opt. Mat. Express*, 9, (2019)
63. R. I. Woodward, et. al. "Tunable Q-switched fiber laser based on saturable edge-state absorption in few-layer molybdenum disulfide (MoS2)," *Opt. Express,* 22, (2014).
64. H. A. Haus, "Parameter ranges for CW passive mode locking," *IEEE J. Quantum. Elec.*, QE-12, 1976.
65. F. X. Kärtner, et. al. "Control of solid state laser dynamics by semiconductor devices," *Optical. Eng*. 34, (1995)
66. L. Li, et. al. Fabrication and characterization of an $As_2Se_3$ optical microwire cladded with perfluorinated CYTOP, *CLEO*, 2016.